\documentclass[journal]{IEEEtran}
\IEEEoverridecommandlockouts
\usepackage{cite}
\usepackage{amsmath,amssymb,amsfonts}
\usepackage{algorithmic}
\usepackage{graphicx}
\usepackage{textcomp}
\usepackage{xcolor}
\usepackage{siunitx}
\usepackage{pgfplots}
\usepackage{tikz}
\usetikzlibrary{quantikz}
\usepackage{braket}
\usepackage{tikz-network}
\usepackage{dblfloatfix}
\usepackage[absolute]{textpos}

\usepackage{amsmath}		

\usepackage{mathptmx}
\usepackage[labelsep=quad,indention=10pt]{subfig}
\usepackage{bm}

\def\BibTeX{{\rm B\kern-.05em{\sc i\kern-.025em b}\kern-.08em
    T\kern-.1667em\lower.7ex\hbox{E}\kern-.125emX}}
\begin{document}

\title{User Trajectory Prediction in Mobile Wireless Networks Using Quantum Reservoir Computing\\
}

\author{Zoubeir~Mlika,~\IEEEmembership{Member,~IEEE,}
        Soumaya~Cherkaoui,~\IEEEmembership{Senior Member,~IEEE,}
        Jean~Frédéric~Laprade, and~Simon~Corbeil-Letourneau,
\thanks{Dr. Mlika was with the Department
of Electrical and Computer Engineering, Université de Sherbrooke, Sherbrooke,
QC, Canada, e-mail: mlika.zoubeir@gmail.com.}
\thanks{Dr. Cherkaoui is with Department of Computer and Software Engineering, Polytechnique Montreal, 
QC, Canada, e-mail: soumaya.cherkaoui@polymtl.ca.}
\thanks{Mr. Laprade is with the Insitut Quantique, Université de Sherbrooke, QC, Canada, e-mail: jean-frederic.laprade@usherbrooke.ca.}
\thanks{Dr. Corbeil-Letourneau is with Thales Digital Solutions, Montreal, QC, Canada, e-mail: simon.corbeil@thalesdigitalsolutions.ca.}}
\thanks{This paper is a preprint of a paper submitted to IET Quantum Communication. If accepted, the copy of record will be available at the IET Digital Library.}

\maketitle

\begin{abstract}
This paper applies a quantum machine learning technique to predict mobile users' trajectories in mobile wireless networks using an approach called quantum reservoir computing (QRC). Mobile users' trajectories prediction belongs to the task of temporal information processing and it is a mobility management problem that is essential for self-organizing and autonomous 6G networks. Our aim is to accurately predict the future positions of mobile users in wireless networks using QRC. To do so, we use a real-world time series dataset to model mobile users' trajectories. The QRC approach has two components: reservoir computing (RC) and quantum computing (QC). In RC, the training is more computational-efficient than the training of simple recurrent neural networks (RNN) since, in RC, only the weights of the output layer are trainable. The internal part of RC is what is called the reservoir. For the RC to perform well, the weights of the reservoir should be chosen carefully to create highly complex and nonlinear dynamics. The QC is used to create such dynamical reservoir that maps the input time series into higher dimensional computational space composed of dynamical states. After obtaining the high-dimensional dynamical states, a simple linear regression is performed to train the output weights and thus the prediction of the mobile users' trajectories can be performed efficiently. In this paper, we apply a QRC approach based on the Hamiltonian time evolution of a quantum system. We simulate the time evolution using IBM gate-based quantum computers and we show in the experimental results that the use of QRC to predict the mobile users' trajectories with only a few qubits is efficient and is able to outperform the classical approaches such as the long short-term memory (LSTM) approach and the echo-state networks (ESN) approach.
\end{abstract}


\section{Introduction}
One of the fundamental theories in physics is quantum mechanics, which is the foundation of all quantum physics such as quantum chemistry, quantum technology, quantum information science. Quantum mechanics describes the physical properties of nature at the scale of atoms and subatomic particles~\cite{feynman2011feynman}. The phenomena of quantum mechanics such as \textit{interference}, \textit{superposition}, and \textit{entanglement} can be exploited to create a new paradigm of computation known as quantum computing~\cite{10.5555/1972505}. Quantum computers are devices that perform quantum computations~\cite{hidary2019quantum} and are shown to have a great potential of fast information processing. For example, the mathematical problem of integer factorization is shown to be solved efficiently on a quantum computer despite it is believed to be intractable on a classical computer~\cite{shor1994algorithms}. 

In this paper, we propose to use quantum mechanics to solve the machine learning task of mobility prediction in mobile wireless networks. This task is nonlinear and belongs to real-world temporal information processing tasks such as time-dependent signal processing, stock-market prediction, natural language processing, etc. The combination of quantum mechanics and machine learning can help solving these real-world temporal information processing tasks. The proposed machine learning approach belongs to the reservoir computing (RC) framework which is inspired by how the brain processes information. Unlike complex recurrent neural networks (RNN) where training is a complex procedure, the RC framework uses the so-called \textit{reservoir} to project the input signals to a higher dimensional space, thus producing a highly dynamical network capable of emulating nonlinear and temporal information processing systems. The reservoir is composed of hidden nodes and visible nodes. The combination of both produces a high-dimensional signal. The training is performed only at the output layer using simple regression analysis~\cite{doi:10.1126/science.1091277,maass2002real,verstraeten2007experimental}. To perfectly emulate nonlinear dynamical systems, the dynamics of the reservoir must involve adequate nonlinearity and memory~\cite{dambre2012information}.

We apply a recently proposed quantum RC (QRC) framework~\cite{fujii2017harnessing} to to the prediction of user trajectory in mobile networks. Mobility management is an important and challenging problem in future 5G/6G networks. In fact, future communication networks will need to provide advanced services for various application areas with diverse requirements in terms of ultra-low latency, data rates,  and massive connectivity. For critical communications and service delivery in particular, issues such as disconnections or service level agreement (SLA) violations can be problematic. A reactive mobility management approach fails to perform well in highly dynamic networks~\cite{li2019deep,filali2020preemptive}. Therefore, a proactive~\cite{filali2020preemptive} mobility management algorithm is required where the mobility patterns of the users can be predicted in advance. 

\subsection{Related Works}
The prediction of user trajectory prediction in wireless networks is not a new topic and has been studied previously in different works~\cite{bahra2020rnn,bahra2021bidirectional,xu2022offloading,dinani2021gossip,ip2021vehicle,liu2019trajectory}. However, to the best of our knowledge, no previous work applied quantum mechanics and machine learning to solve the challenging problem of trajectory prediction in mobile wireless networks. Quantum mechanics and machine learning have been proposed to solve machine learning tasks such as classification and prediction~\cite{kutvonen2020optimizing,feng2020monthly,araujo2010quantum,ahmed2022quantum,suzuki2022natural,fujii2017harnessing} but they were not applied to a practical mobile wireless network problem.

On the one hand, in~\cite{bahra2020rnn} the authors proposed a mobility prediction approach based on RNN. Precisely, the authors proposed to use a gated recurrent unit (GRU) and an long short-term memory (LSTM) techniques to predict the global positioning system (GPS) coordinates of a mobile user; taken from the Geolife dataset~\cite{10.1145/1658373.1658374}. Their approaches were initialized with a pre-processing step to reduce the number of points, and thus the prediction complexity, in the time series data. In~\cite{bahra2021bidirectional}, the same authors proposed the same mobility prediction approaches but using bidirectional analysis and three different datasets: Geolife, open street map, and T-drive trajectory. In~\cite{xu2022offloading} the authors studied two data-driven mobility prediction algorithms in vehicular networks and measured how inaccurate their prediction accuracy was. To overcome the imperfect mobility prediction problem, the authors designed a new scheduling system to offload cellular traffic to vehicular network and showed that prediction accuracy was improved. In~\cite{dinani2021gossip}, the authors proposed a decentralized traffic management scheme to predict vehicle trajectory. Each vehicle predicted its own trajectory by training an LSTM model. The proposed approach used gossip learning and iterative model averaging is used to build a global model. The evaluation was based on urban vehicular network environment where the dataset of the Luxembourg SUMO Traffic scenario~\cite{7385539} was used. In~\cite{ip2021vehicle}, the authors proposed an RNN-based LSTM approach to predict vehicle trajectory using a taxi dataset of 442 taxis running in Porto, Portugal. The results showed the effectiveness of the prediction with a prediction performance higher than 89\%. The improvement was observed especially when more data are available prior to the next prediction (prediction of the next cell). In~\cite{liu2019trajectory}, the authors studied the problem of power control and trajectory planning in unmanned aerial vehicle networks. The aim was to maximize the sum rate while guaranteeing the rate requirement of the users. First, an UAV placement algorithm was proposed based on Q-learning to position the UAVs. Then, using the real data of the users collected from Twitter, the authors proposed an echo-state network (ESN) to predict the future position of the users. Finally, a Q-learning algorithm was proposed to predict the UAV positions in future timesteps. The results showed that as the size of the reservoir in ESN increases, the prediction increases and a sum rate gain of 17\% was obtained.

On the other hand, in~\cite{kutvonen2020optimizing}, the authors proposed a QRC system where the dynamics of the reservoir evolve according to Hamiltonian evolution in the fully connected transverse field Ising (FC-TFI) model. The authors studied the memory capacity and the accuracy of the proposed QRC by varying different parameters such as the inter-spin interactions of the FC-TFI model and the time evolution scale. They showed the existence of an optimal time evolution scale at which the capacity of the QRC is maximized. Finally, they applied their method on stock prediction data. In~\cite{araujo2010quantum}, the authors aimed to overcome the random walk dilemma for financial time series prediction using a quantum-inspired hybrid method. Their method used a qubit multilayer perceptron (QuMLP) and a quantum-inspired evolutionary algorithm (QIEA). The QIEA was a search-based algorithm that trains the QuMLP to determine parameters like the maximum number of time lags to represent the financial time series, the number of units in the QuMLP hidden layer. Four time series datasets were evaluated that correspond to the daily records of Nasdaq stock market and the results were shown to be superior compared to classical MLP. In~\cite{suzuki2022natural}, the authors proposed to use superconducting quantum computing devices as the reservoir in a QRC system. They showed that the inherent noise characterizing nowadays quantum computers is advantageous and helped in providing dissipative dynamics capable to learn a dynamical system and solve a temporal information processing task. The authors studied the prediction of the nonlinear autoregressive moving average (NARMA) time series as well as an experimental-based classification problem to classify three objects: one LEGO cube, one polylactic acid (PLA) cube and PLA sphere. The time series for the classification problem were generated by the triboelectric nanogenerator (TENG) sensor of a robotic gripper grabbing the objects. In both the prediction and the classification problems, the proposed gate-based QRC approach showed a higher performance than classical linear regression or classification models. The authors concluded that a noisy quantum device can potentially work as a reservoir computer, and notably, that the undesirable quantum noise can be used as a rich computation resource for machine learning tasks. In~\cite{fujii2017harnessing}, the authors proposed to exploit the natural quantum dynamics of ensemble systems to solve a temporal information processing task using QRC. The proposed QRC framework was shown to enable ensemble quantum systems to universally emulate nonlinear dynamical systems including classical chaos. The authors performed a number of numerical experiments for quantum systems with 5–7 qubits and showed their superiority in terms of computational capabilities compared to conventional RNNs composed of 100–500 nodes.

\subsection{Paper Organisation}
The rest of the paper is organized as follows: Section~\ref{qrc} first introduces the basic concepts of quantum mechanics and then describes the QRC framework used to predict user trajectory. Section~\ref{bench} describes classical state-of-the-art solutions. Section~\ref{eval} describes the real-world experimental setups and illustrates important results and conclusions. Finally, Section~\ref{cl} draws some important conclusions.

\section{Quantum Reservoir Computing Approach}\label{qrc}
The idea of the proposed QRC framework is essentially based on applying the FC-TFI model introduced in~\cite{fujii2017harnessing,kutvonen2020optimizing}. In this framework, the dynamics of the recurrent part of the network (the reservoir) are governed by the dynamics of rich quantum mechanical systems. Before describing in depth the QRC framework, we start by introducing some basic concepts about quantum computation and quantum information. 

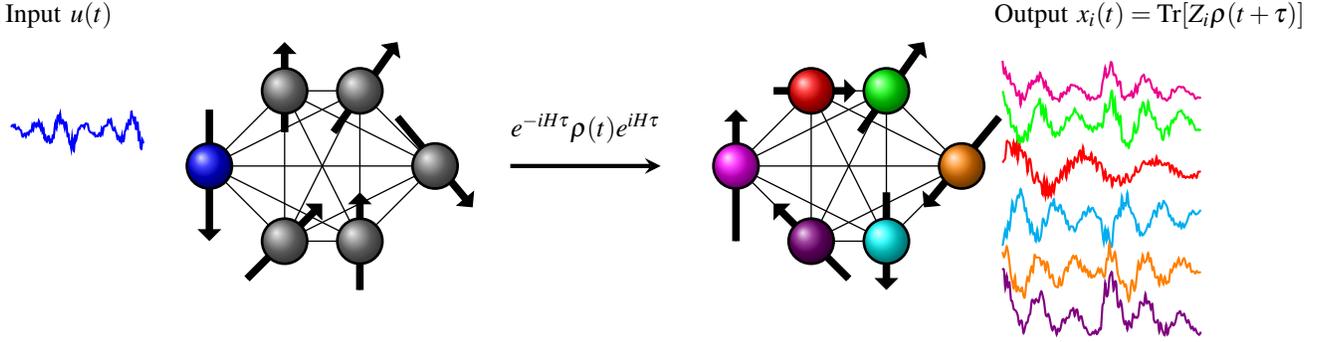
\begin{figure*}
    \centering
   \begin{tikzpicture}[->]
        \draw[-,x=-.5,y=2,xscale=1,yscale=3,
            declare function={
            excitation(\t,\w) = sin(\t*\w);
            noise = rnd - 0.5;
            source(\t) = excitation(\t,20) + noise;
            filter(\t) = 1 - abs(sin(mod(\t, 50)));
            speech(\t) = 1 + source(\t)*filter(\t);},
            blue, thick, smooth,          
            domain=50:150, samples=144,      
            ] plot (\x,{6+speech(\x)}); 
        \draw[-,x=15,y=2,xscale=0.05,yscale=3,
            declare function={
            excitation(\t,\w) = sin(\t*\w);
            noise = rnd + 0.5;
            source(\t) = excitation(\t,20) + noise;
            filter(\t) = 1 - abs(sin(mod(\t, 50)));
            speech(\t) = 1 + source(\t)*filter(\t);},
            magenta, thick, smooth,          
            domain=400:500, samples=144,      
            ] plot (\x,{8+speech(\x)}); 
        \draw[-,x=15,y=3,xscale=0.05,yscale=3,
            declare function={
            excitation(\t,\w) = sin(\t*\w);
            noise = rnd - 0.5;
            source(\t) = excitation(\t,20) + noise;
            filter(\t) = 1 - abs(sin(mod(\t, 50)));
            speech(\t) = 1 + source(\t)*filter(\t);},
            green, thick, smooth,          
            domain=400:500, samples=144,      
        ] plot (\x,{4+speech(\x)}); 
        \draw[-,x=15,y=3,xscale=0.05,yscale=3,
            declare function={
            excitation(\t,\w) = sin(\t*\w);
            noise = rnd - 0.5;
            source(\t) = excitation(\t,20) + noise;
            filter(\t) = 1 - abs(sin(mod(\t, 50)));
            speech(\t) = 1 + source(\t)*filter(\t);
            speech2(\t) = speech(\t/2);},
            red, thick, smooth,          
            domain=400:500, samples=144,      
        ] plot (\x,{2+speech2(\x)}); 
        \draw[-,x=15,y=3,xscale=0.05,yscale=3,
            declare function={
            excitation(\t,\w) = sin(\t*\w);
            noise = rnd - 0.5;
            source(\t) = excitation(\t,20) + noise;
            filter(\t) = 1 - abs(sin(mod(\t, 50)));
            speech(\t) = 1 + source(\t)*filter(\t);
            speech3(\t) = speech(\t/4);},
            cyan, thick, smooth,          
            domain=400:500, samples=144,      
        ] plot (\x,{0+speech(\x+100)}); 
        \draw[-,x=15,y=3,xscale=0.05,yscale=3,
            declare function={
            excitation(\t,\w) = sin(\t*\w);
            noise = rnd - 1;
            source(\t) = excitation(\t,20) + noise;
            filter(\t) = 1 - abs(sin(mod(\t, 50)));
            speech(\t) = 1 + source(\t)*filter(\t);
            speech3(\t) = speech(\t/4);},
            orange, thick, smooth,          
            domain=400:500, samples=144,      
        ] plot (\x,{-2+speech(\x)}); 
        \draw[-,x=15,y=3,xscale=0.05,yscale=3,
            declare function={
            excitation(\t,\w) = sin(\t*\w);
            noise = rnd + 1;
            source(\t) = excitation(\t,20) + noise;
            filter(\t) = 1 - abs(sin(mod(\t, 50)));
            speech(\t) = 1 + source(\t)*filter(\t);
            speech3(\t) = speech(\t/4);},
            violet, thick, smooth,          
            domain=400:500, samples=144,      
        ] plot (\x,{-5+speech(\x)}); 
        \node at (-2,3) {Input $u(t)$};
        \node at (12.5,3) {Output $x_i(t)=\text{Tr}[Z_i\rho(t+\tau)]$};
        \draw[-triangle 90,line width=0.2mm,
        postaction={draw, line width=1mm, shorten >=1mm, -}] (0.5,-0.5) -- (1.5,0.5);
        \draw[ultra thick,-triangle 90,line width=0.2mm,
        postaction={draw, line width=1mm, shorten >=1mm, -}] (2,-0.65) -- (2,0.65);
        \draw[ultra thick,-triangle 90,line width=0.2mm,
        postaction={draw, line width=1mm, shorten >=1mm, -}] (2.5,1.65) -- (3.5,0.45);
        \draw[ultra thick,-triangle 90,line width=0.2mm,
        postaction={draw, line width=1mm, shorten >=1mm, -}] (1.65,1.45) -- (2.5,2.65);
        \draw[ultra thick,-triangle 90,line width=0.2mm,
        postaction={draw, line width=1mm, shorten >=1mm, -}] (0,1.75) -- (0,0);
        \draw[ultra thick,-triangle 90,line width=0.2mm,
        postaction={draw, line width=1mm, shorten >=1mm, -}] (1,1.45) -- (1,2.65);
        \Vertex[style={shading=ball,ball color=gray},x=2,y=0]{A}
        \Vertex[style={shading=ball,ball color=gray},x=1,y=0]{B}
        \Vertex[style={shading=ball,ball color=blue},x=0,y=1]{C}
        \Vertex[style={shading=ball,ball color=gray},x=1,y=2]{D}
        \Vertex[style={shading=ball,ball color=gray},x=2,y=2]{E}
        \Vertex[style={shading=ball,ball color=gray},x=3,y=1]{F}
        \Edge[lw=.5,color=black](A)(B)
        \Edge[lw=.5,color=black](A)(C)
        \Edge[lw=.5,color=black](A)(D)
        \Edge[lw=.5,color=black](A)(E)
        \Edge[lw=.5,color=black](A)(F)
        \Edge[lw=.5,color=black](B)(C)
        \Edge[lw=.5,color=black](B)(D)
        \Edge[lw=.5,color=black](B)(E)
        \Edge[lw=.5,color=black](B)(F)
        \Edge[lw=.5,color=black](C)(D)
        \Edge[lw=.5,color=black](C)(E)
        \Edge[lw=.5,color=black](C)(F)
        \Edge[lw=.5,color=black](D)(E)
        \Edge[lw=.5,color=black](D)(F)
        \Edge[lw=.5,color=black](E)(F)
        \draw[>=stealth,ultra thick] (4,1) -- (6,1);
        \node at (5,1.5) {$e^{-iH\tau}\rho(t)e^{iH\tau}$};
        \draw[-triangle 90,line width=0.2mm,
        postaction={draw, line width=1mm, shorten >=1mm, -}] (8.5,-0.5) -- (7.5,0.5);
        \draw[ultra thick,-triangle 90,line width=0.2mm,
        postaction={draw, line width=1mm, shorten >=1mm, -}] (9,0.65) -- (9,-0.65);
        \draw[ultra thick,-triangle 90,line width=0.2mm,
        postaction={draw, line width=1mm, shorten >=1mm, -}] (10.5,1.65) -- (9.5,0.45);
        \draw[ultra thick,-triangle 90,line width=0.2mm,
        postaction={draw, line width=1mm, shorten >=1mm, -}] (8.65,1.45) -- (9.5,2.65);
        \draw[ultra thick,-triangle 90,line width=0.2mm,
        postaction={draw, line width=1mm, shorten >=1mm, -}] (7,0) -- (7,1.75);
        \draw[ultra thick,-triangle 90,line width=0.2mm,
        postaction={draw, line width=1mm, shorten >=1mm, -}] (7.5,2) -- (8.6,2);
        \Vertex[style={shading=ball,ball color=cyan},x=9,y=0]{A}
        \Vertex[style={shading=ball,ball color=violet},x=8,y=0]{B}
        \Vertex[style={shading=ball,ball color=magenta},x=7,y=1]{C}
        \Vertex[style={shading=ball,ball color=red},x=8,y=2]{D}
        \Vertex[style={shading=ball,ball color=green},x=9,y=2]{E}
        \Vertex[style={shading=ball,ball color=orange},x=10,y=1]{F}
        \Edge[lw=.5,color=black](A)(B)
        \Edge[lw=.5,color=black](A)(C)
        \Edge[lw=.5,color=black](A)(D)
        \Edge[lw=.5,color=black](A)(E)
        \Edge[lw=.5,color=black](A)(F)
        \Edge[lw=.5,color=black](B)(C)
        \Edge[lw=.5,color=black](B)(D)
        \Edge[lw=.5,color=black](B)(E)
        \Edge[lw=.5,color=black](B)(F)
        \Edge[lw=.5,color=black](C)(D)
        \Edge[lw=.5,color=black](C)(E)
        \Edge[lw=.5,color=black](C)(F)
        \Edge[lw=.5,color=black](D)(E)
        \Edge[lw=.5,color=black](D)(F)
        \Edge[lw=.5,color=black](E)(F)
    \end{tikzpicture}%
    \caption{\footnotesize An example of a QR with 6 qubits shown with gray balls (one qubit is in blue color to indicate the input qubit). Each qubit's state is given by the corresponding arrow. An input signal (a time series) of length $T$ is injected into one qubit at some initial time $t$ and the system evolves for a time period $\tau$ according to the dynamics of the QR. After a time $\tau$, a measurement is performed. By repeating this process multiple times for the same input signal, we obtain the average spin values in the z-direction of each spin (qubit). Note that if the input signal has length $T$ and we inject into the reservoir the values of the signal one at a time at each time step $\tau$, then the simulation will take a total of $T\tau$ time.}
    \label{fig:qrc}
\end{figure*}

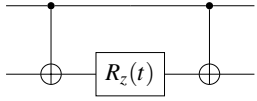
\begin{figure}[!h]
\centering
\resizebox{0.2\textwidth}{!}{\begin{quantikz}[thin lines,row sep={1cm,between origins},transparent]
\qw & \ctrl{1} & \qw & \ctrl{1} & \qw\\
\qw & \targ{} & \gate{R_z(t)} & \targ{} & \qw
\end{quantikz}}
\caption{\footnotesize Simulation of the time evolution of the Hamiltonian $ZZ$, i.e., $e^{-iZZt}$.}
\label{fig:ZZ}
\end{figure}

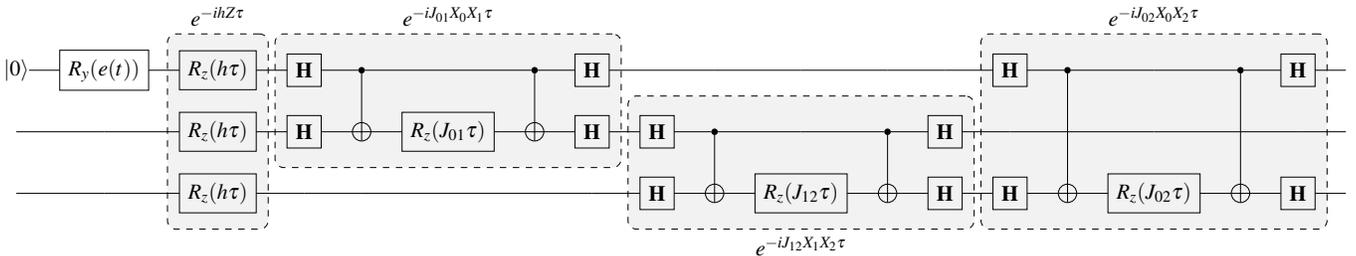
\begin{figure*}[!b]
\centering
\resizebox{\textwidth}{!}{\begin{quantikz}[thin lines,row sep={1cm,between origins},transparent]
\ket{0} & \gate{R_y(e(t))} & \gate{R_z(h\tau)}\gategroup[3,steps=1,style={dashed, rounded corners,fill=gray!10, inner xsep=2pt}, background]{{$e^{-ihZ\tau }$}} & \gate{\mathbf{H}}\gategroup[2,steps=5,style={dashed, rounded corners,fill=gray!10, inner xsep=2pt}, background]{{$e^{-iJ_{01}X_0X_1\tau}$}} & \ctrl{1} & \qw & \ctrl{1} & \gate{\mathbf{H}} & \qw & \qw & \qw & \qw & \qw & \gate{\mathbf{H}}\gategroup[3,steps=5,style={dashed, rounded corners,fill=gray!10, inner xsep=2pt}, background]{{$e^{-iJ_{02}X_0X_2\tau}$}} & \ctrl{1} & \qw & \ctrl{1} & \gate{\mathbf{H}} & \qw\\
\qw & \qw & \gate{R_z(h\tau)} & \gate{\mathbf{H}} & \targ{} & \gate{R_z(J_{01}\tau)} & \targ{} & \gate{\mathbf{H}} & \gate{\mathbf{H}}\gategroup[2,steps=5,style={dashed, rounded corners,fill=gray!10, inner xsep=2pt}, background, label style={label position=below,anchor= north,yshift=-0.2cm}]{{$e^{-iJ_{12}X_1X_2\tau}$}} & \ctrl{1} & \qw & \ctrl{1} & \gate{\mathbf{H}} & \qw  & \qw \vqw{1} & \qw & \qw \vqw{1} & \qw & \qw \\
\qw & \qw & \gate{R_z(h\tau)} & \qw & \qw & \qw & \qw & \qw & \gate{\mathbf{H}} & \targ{} & \gate{R_z(J_{12}\tau)} & \targ{} & \gate{\mathbf{H}} & \gate{\mathbf{H}} & \targ{} & \gate{R_z(J_{02}\tau)} & \targ{} & \gate{\mathbf{H}} & \qw
\end{quantikz}}
\caption{\footnotesize Quantum circuit realization to calculate the exponential of the FC-TFI Hamiltonian with 3 qubits. The input is injected onto the first qubit which is prepared in the state $|0\rangle$. Then, the time series data $u(t)$ is encoded in the first qubit using the single-qubit rotation gate about the $Y$-axis, $R_y(e(t))$. Here, $e(t)$ is chosen such that the state of the first qubit will be $\sqrt{1-u(t)}|0\rangle+\sqrt{u(t)}|1\rangle$ just after the gate $R_y(e(t))$, i.e., $e(t)=2\arcsin\sqrt{u(t)}$. We assume a full coupling between the qubits so that each qubit interacts with all other qubits. The resulting quantum circuit can be simplified but is kept as is for clarity.}
\label{fig:quantum:circuit}
\end{figure*}

\subsection{Qubit State}
Contrary to the classical bit---the minimum unit of information in classical computing, the quantum bit (or simply \textit{qubit}) is the minimum unit of quantum information in quantum computing. A qubit is a two-state quantum-mechanical system such as the spin of the electron with the two states of spin-up and spin-down. As opposed to a classical system in which a bit is only in one state at a time, a qubit, based on the laws of quantum mechanics, can be simultaneously in a coherent superposition of the two states. Mathematically, a qubit is a vector in a two-dimensional complex vector space spanned by the orthonormal basis $\{\vec{0},\vec{1}\}$. In quantum mechanics, the vectors $\vec{0}$ and $\vec{1}$ of the basis $\{\vec{0},\vec{1}\}$ are conventionally written in the Dirac notation (or the bra-ket notation) as $|0\rangle$ and $|1\rangle$, respectively. A pure qubit state can be in any coherent superposition (linear combinations) of the two basis states $|0\rangle$ and $|1\rangle$, i.e., a qubit can be written as~\cite{10.5555/1972505}:
\begin{align}
    |\psi\rangle=\alpha|0\rangle+\beta|1\rangle,
\end{align}
where $\alpha$ and $\beta$ are complex numbers that are called the probability amplitudes. When a qubit is measured, analogously, when a bit is read, the outcome of the measurement is state $|0\rangle$ with probability $|\alpha|^2$, and state $|1\rangle$ with probability $|\beta|^2$. Since the probabilities must sum to $1$, the coefficients obey to $|\alpha|^2+|\beta|^2=1$. In other words, the qubit's state must be normalized to length $1$. In summary, a qubit can be in a continuum of states between $|0\rangle$ and $|1\rangle$ until it is \textit{measured}.

An $n$-qubit quantum-mechanical system is described by the tensor product space of $n$ two-dimensional complex vector spaces. That is, an $n$-qubit quantum-mechanical system has $2^n$ computational basis states $|x_1x_2\ldots x_n\rangle$ or equivalently $|x_1\rangle\otimes|x_2\rangle\otimes\cdots\otimes|x_n\rangle$ where the notation $\otimes$ denotes the tensor product and $x_i\in\{0,1\}$ for all $i\in\{0,1,\ldots,n-1\}$. Another way to characterize a quantum-mechanical system of $n$ qubits is with the density operator, denoted as $\rho$, which is a Hermitian matrix of size $2^n\times2^n$. The density matrix can be seen as a statistical mixture of pure states of the quantum-mechanical system~\cite{10.5555/1972505}.

Given the state $\ket{\psi}$ describing a quantum mechanical system, how does it change with time? The second postulate of quantum mechanics answers this question. It states that the evolution of a closed quantum system is described by a unitary transformation. Mathematically, if the quantum state is $|\psi(t_1)\rangle$ at time $t_1$, then at a later time $t_2$, the state of the quantum system is given by the following relation~\cite{10.5555/1972505}:
\begin{align}\label{postulate2}
    |\psi(t_2)\rangle=U(t_1,t_2)|\psi(t_1)\rangle,
\end{align}
where $U(t_1,t_2)$ is a unitary operator that depends only on $t_1$ and $t_2$. Based on the Schr\"odinger’s equation and the second postulate of quantum mechanics, we can obtain the following equation for the unitary $U(t_1,t_2)$:
\begin{align}
    U(t_1,t_2)=\exp\bigl(-iH(t_2-t_1)/\hbar\bigr),
\end{align}
where $i$ denotes the imaginary unit and $H$ is a fixed Hermitian operator known as the Hamiltonian (a Hermitian matrix of size $2^n\times2^n$) of the closed quantum-mechanical system and $\hbar$ is Planck’s constant.

In general, for a closed quantum-mechanical system, the time evolution is given by a unitary operator $U=\exp(-iHt)$ for some Hermitian operator $H$ (where the constant $\hbar$ is absorbed in the Hamiltonian $H$). If the system is described in terms of the density matrix $\rho(t)$ at time $t$, then the time evolution for a time interval $\tau$ is given by~\cite{10.5555/1972505}:
\begin{align}\label{shro1}
    \rho(t+\tau)=e^{-iH\tau}\rho(t)e^{iH\tau}.
\end{align}

\subsection{Qubit Measurement}
As discussed previously, a closed quantum-mechanical system evolves according to unitary evolution. When an external physical system interacts with this closed system to perform a measurement, the system is no longer closed. The third postulate of quantum mechanics describes the effect of a measurement on a quantum mechanical system by a set of projective operators $\{P_m\}$ such that $\sum_mP_m=I$ and $P_mP_n=\delta_{mn}P_m$ (where $\delta_{mn}$ is the Kronecker delta function). The probability to obtain the measurement outcome $m$ for the state $\rho$ is given by $p(m)=\text{Tr}\bigl[P_m\rho\big]$ where $\text{Tr}[\cdot]$ denotes the trace of a matrix. The state after the measurement is given by $P_m\rho P_m/p(m)$, that is, the projective measurement modifies the quantum-mechanical system. By repeating the projective measurements, after preparing the quantum state again, we can calculate average values $\langle O\rangle=\text{Tr}\bigl[O\rho\bigr]$ of an observable given by $O=\sum_ma(m)P_m$ according to its spectral decomposition, where $a(m),\,\forall m$ are the eigenvalues of $O$~\cite{10.5555/1972505}.

\subsection{Quantum Reservoir Dynamics}
The nodes of the quantum reservoir network are given by the orthogonal basis of the quantum states. This means that for $n$ qubits, we have $2^n$ basis states. To be efficient, quantum reservoir computing requires complex and bounded dynamics and large network sizes~\cite{kutvonen2020optimizing}. It is known that these requirements can be naturally met by interacting quantum-mechanical spin systems~\cite{kutvonen2020optimizing}. These spin systems have (i) state space that scales exponentially with the number of spins and (ii) complex dynamics governed by unitary operators. 

In this paper, we consider the extensively studied model called the fully connected transverse field Ising (FC-TFI) model~\cite{fujii2017harnessing,kutvonen2020optimizing}. The Hamiltonian of this model is given by:
\begin{align}\label{ising}
    H = \sum_{i,j}J_{i,j}X_iX_j+h_iZ_i,
\end{align}
where $X_i$ and $Z_i$ are the Pauli $X$ and $Z$ operators at qubit $i$, the coefficient $h_i$ denotes the coupling to an external magnetic field of qubit $i$ and the coefficient $J_{ij}$ denotes the inter-qubit interactions. \if A qubit's state in the FC-TFI model is given by the state of a qubit in a two-dimensional complex vector spanned by the orthonormal basis $\{|0\rangle, |1\rangle\}$ of the Pauli $Z$ operator.\fi In the FC-TFI model, all the qubits, as shown in~\eqref{ising}, interact with each other in the x-direction and are coupled to an external magnetic field in the z-direction.

In the QRC framework, an input sequence of length $T$, $u(t) \in \{u_0, u_1,\ldots, u_{T-1}\}$ where $u_i \in [0,1]$ is given by the trajectory of a mobile user (i) is injected into the quantum reservoir network, (ii) evolves according the quantum dynamics, and then (iii) is extracted as output to be analyzed. The idea of the QRC framework~\cite{kutvonen2020optimizing,fujii2017harnessing} is to inject the input at time $t$, given by $u(t)$, through the first qubit by setting the state of this qubit to 
\begin{align}
    |\psi_{u(t)}\rangle=\sqrt{1-u(t)}|0\rangle+\sqrt{u(t)}|1\rangle, \;u(t)\in[0,1].
\end{align}
The density matrix of the system is then given by:
\begin{align}
    \rho=|\psi_{u(t)}\rangle\langle\psi_{u(t)}|\otimes\text{Tr}_1\bigl[\rho\bigr],
\end{align}
where $\text{Tr}_1\bigl[\rho\bigr]$ denotes the partial trace operator over the first qubit. In other words, the quantum state of the system is given by the tensor product of the first qubit's state $|\psi_{u(t)}\rangle\langle\psi_{u(t)}|$ and the remaining qubits' states $\text{Tr}_1\bigl[\rho\bigr]$. After injecting the input $u(t)$ into the first qubit, the quantum-mechanical system continues evolving for a time $\tau$ according the Shr\"odinger's equation as discussed previously in~\eqref{postulate2} and~\eqref{shro1}. The dynamics of the quantum reservoir is governed by~\eqref{shro1} during the time $\tau$ and the information that was encoded in the first qubit will spread through the quantum reservoir. An illustrative example of the QRC is given in Fig.~\ref{fig:qrc}.
It is clear from Fig.~\ref{fig:qrc} that the QRC approach allows to project the input signal (the time series corresponding to the mobile user trajectory) into a high-dimensional space and, from the dynamics of the reservoir, extract outputs corresponding to the evolution of the Pauli-Z operator's average values of the selected qubits. Note that if the input signal has length $T$ and we inject into the reservoir the values of the signal one at a time at each time step $\tau$, then the simulation will take a total of $T\tau$ time. 

To let the system evolve for a time $\tau$, one must apply the evolution operator according to $e^{-iH\tau}\rho(t)e^{iH\tau}$. On a gate-based quantum computer, this can be achieved by applying the Suzuki-Trotter decomposition~\cite{10.5555/1972505} which states that the evolution of an operator consisting in a sum of local operators for a time $\tau$ can be expressed as the evolution for a time ($\tau/\kappa$) of a product of local operators, repeated $\kappa$ times. In the limit of $\kappa \rightarrow \infty$, this formula is exact. More formally, if $H = \sum_ih_i$, then
$e^{-iH\tau} = e^{-i\sum_ih_i\tau} = \lim_{\kappa \rightarrow \infty} \big(\prod_i e^{-ih_i\tau/\kappa} \big)^\kappa$
For finite values of $\kappa$, the error made by this approximation is proportional to $\tau^2/\kappa$.
\if This calculation involves a complex task consisting in evaluating the exponential of a matrix (the Hamiltonian).\fi Since the Hamiltonian of our QRC system is based on the FC-TFI model, our task consists in implementing the evolution of the two qubits interaction operator $e^{-iJ_{ij}X_iX_j\tau/\kappa}$ and the single qubit operator $e^{-ih_iZ_i\tau/\kappa}$.
With the rotation gates defined by $R_{\{X,Z\}}(\theta) = e^{-i\frac{\theta}{2}\{X,Z\}}$, we obtain
\begin{align}
    e^{-ih_iZ_i\tau/\kappa} = R_z(2h_i\tau/\kappa).
\end{align}

To obtain the evolution of the interaction term $e^{-iJ_{ij}X_iX_j\tau/\kappa}$, we note that the unitary operator $e^{-i\frac{\theta}{2}ZZ}$ is realized by the sequence of gates shown in Fig.~\ref{fig:ZZ} and that we have the identity $X = \mathbf{H}Z\mathbf{H}$ where $\mathbf{H}$ corresponds to the Hadamard gate. Thus, $e^{-iJ_{ij}X_iX_j\tau/\kappa}$ is realized, as shown in Fig.~\ref{fig:quantum:circuit}, with four Hadamard gates and the block given in Fig.~\ref{fig:ZZ}.


Fig.~\ref{fig:quantum:circuit} illustrates the basic idea of how to simulate the time evolution of the Hamiltonian given in~\eqref{ising} with full coupling between three qubits. Given the Hamiltonian $H$, which is given by $h\bigl(Z_0+Z_1+Z_2\bigr)+J_{0,1}X_0X_1+J_{1,2}X_1X_2$ for $3$ qubits, we can use IBM gate-based quantum computers and implement the quantum circuit given in Fig.~\ref{fig:quantum:circuit} to estimate the time evolution of the Hamiltonian.

The objective of the mobility trajectory prediction problem is to find, using the QRC system and given the trajectory of a mobile user $u(t)$, a nonlinear function $f(u(t))$ such that the prediction error (e.g., mean-squared error) between $f(u(t))$ and a target output $\bar{u}(t)$ (teacher output) is minimized. The role of the QRC system is to emulate this nonlinear function and thus produce a highly complex and nonlinear system capable of predicting temporal input data. The output signal, $x_i(t)$, obtained after measuring qubit $i$ at each time $t$ is used at the output layer of the QRC system and combined with the readout weights $\mathbf{W}^{\text{out}}$ to perform the prediction. This is done using a simple linear regression model. Let $\mathbf{x}(t)$ denotes the vector $[x_1(t),\ldots,x_n(t)]^\top$. After collecting enough pairs of input-output signals $(u(t),\mathbf{x}(t))$ for each time $t$, we use a linear regression technique to train the readout weight $\mathbf{W}^{\text{out}}$ of size $n+1$ where $n$ is the number of qubits used and the additional term is used for the bias. This can be done simply using the Moore–Penrose pseudo inverse, i.e., we need to solve the following linear system:
\begin{align}
    \mathbf{\bar{u}}=\mathbf{\bar{X}}\mathbf{W}^\text{out},
\end{align}
where $\mathbf{\bar{u}}$ is the target signal of size $T$, the matrix $\mathbf{\bar{X}}$ denotes the modified output signals of the QR after adding the bias of size $T\times(n+1)$.

 

\section{Classical Benchmark Approaches}\label{bench}

\subsection{LSTM Prediction}
To compare the QRC approach, we implemented the well-known RNN-based approach of LSTM~\cite{yu2019review}. RNN suffers from the vanishing/exploding gradient problem in which long-term gradients that are back-propagated can either vanish (reach zero) or explode (reach infinity). LSTM was proposed to deal particularly with the vanishing gradient problem. In the LSTM approach, a RNN is able to keep track of long-term dependencies in the input sequences thanks to the use of feedback connections and special neural network units (which allow information to persist). In LSTM, there is a key component called the cell state to which information are added or removed using other components called gates (e.g., forget gate layer, input gate layer, etc.). Other details about the proposed LSTM network are described in Section~\ref{eval}.

\subsection{ESN Prediction}
We compared the QRC approach to its classical counterpart, the echo-state network (ESN)~\cite{9127499}. In the ESN approach, a RNN is also used but in the context of RC. The difference between RNN and RC is that in the former the hidden layers weights are optimized during training but in the latter, these weights are randomly generated and thus are fixed during training. In ESN, only the weights of the output layers are trainable. The structure of the reservoir (i.e., the hidden layers) should be constructed to generate highly complex dynamics so that the training of the output weights reproduce with accuracy specific temporal patterns. An input signal will be injected into the reservoir that will generate a high-dimensional signal using its internal reservoir states (its dynamics). Next, the high-dimension signal is passed to the output layer for training. Other details about the proposed ESN are also given in Section~\ref{eval}.

\section{Experimental Evaluation}\label{eval}
In this section, we evaluate experimentally using a real-world dataset the performance of the QRC and compare it to the LSTM and ESN approaches. We use a GPS dataset called the Geolife dataset~\cite{10.1145/1658373.1658374, bahra2020rnn} that contains GPS trajectories of 182 mobile users, i.e., it contains the longitude and latitude of each mobile user. It was collected between the year 2007 and the year 2012; over a period of almost five years. Each mobile user in the Geolife dataset has its own trajectory, meaning that the length of each trajectory (the number of data of available points) varies from one user to another. Our objective is to accurately predict future mobile user locations. We have randomly chosen three users from this dataset, which are indexed as user \#1, user \#153, and user \#175 and we used their trajectories as movement history to learn their mobility behaviours using the QRC approach and compared to the LSTM and ESN approaches. Note that the LSTM and ESN approaches are not tuned to optimality. We only perform a grid search method to select good LSTM and ESN parameters. The aim is to illustrate how well can the QRC approach perform compared to classical approaches.

\begin{figure}[t!]
    \centering
       \includegraphics[scale=1]{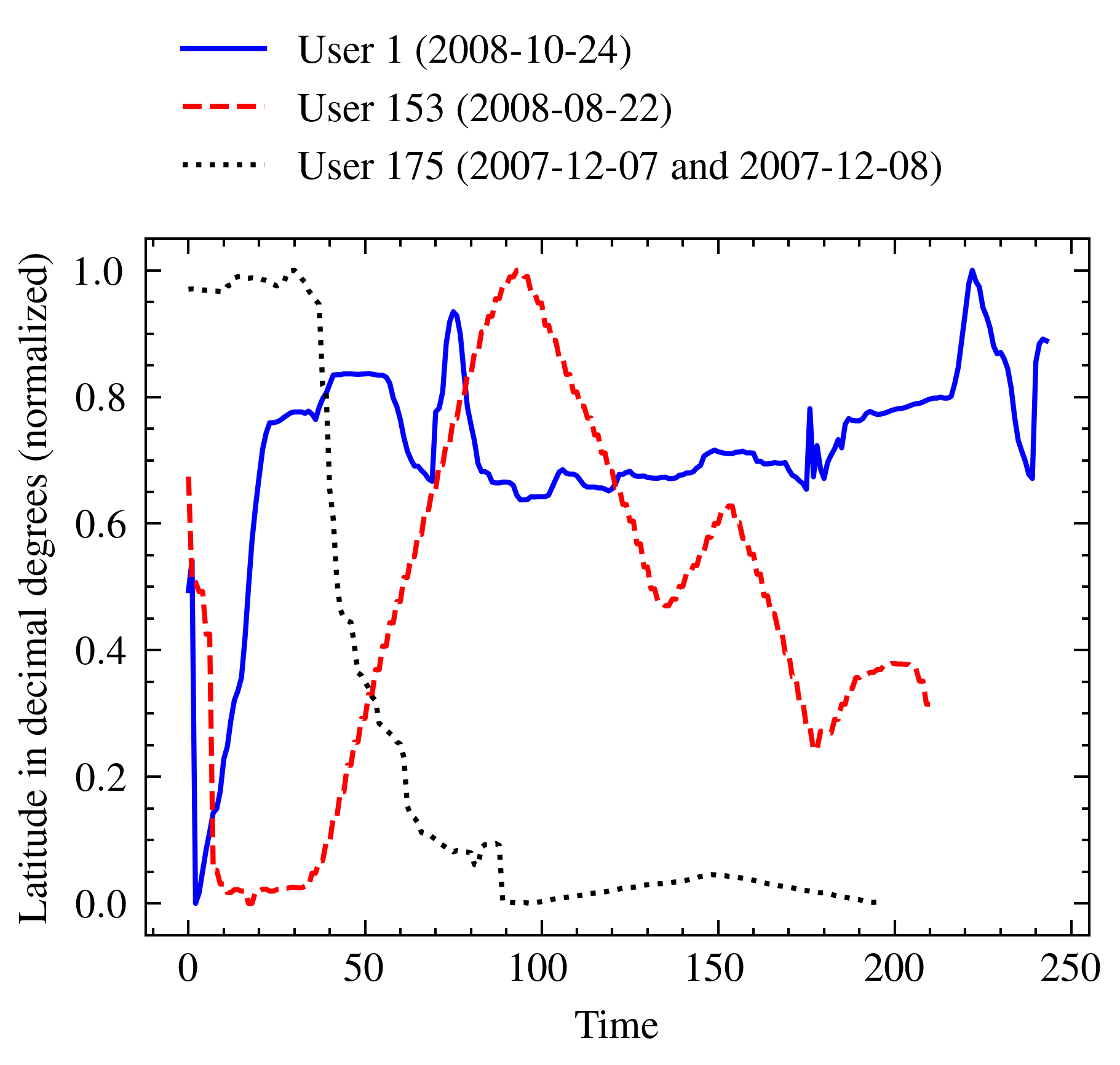}
       \caption{\footnotesize Overview of the latitude variable for the three users.}
       \label{fig:lat}
\end{figure}

\begin{figure}[!b]
    \centering
       \includegraphics[scale=1]{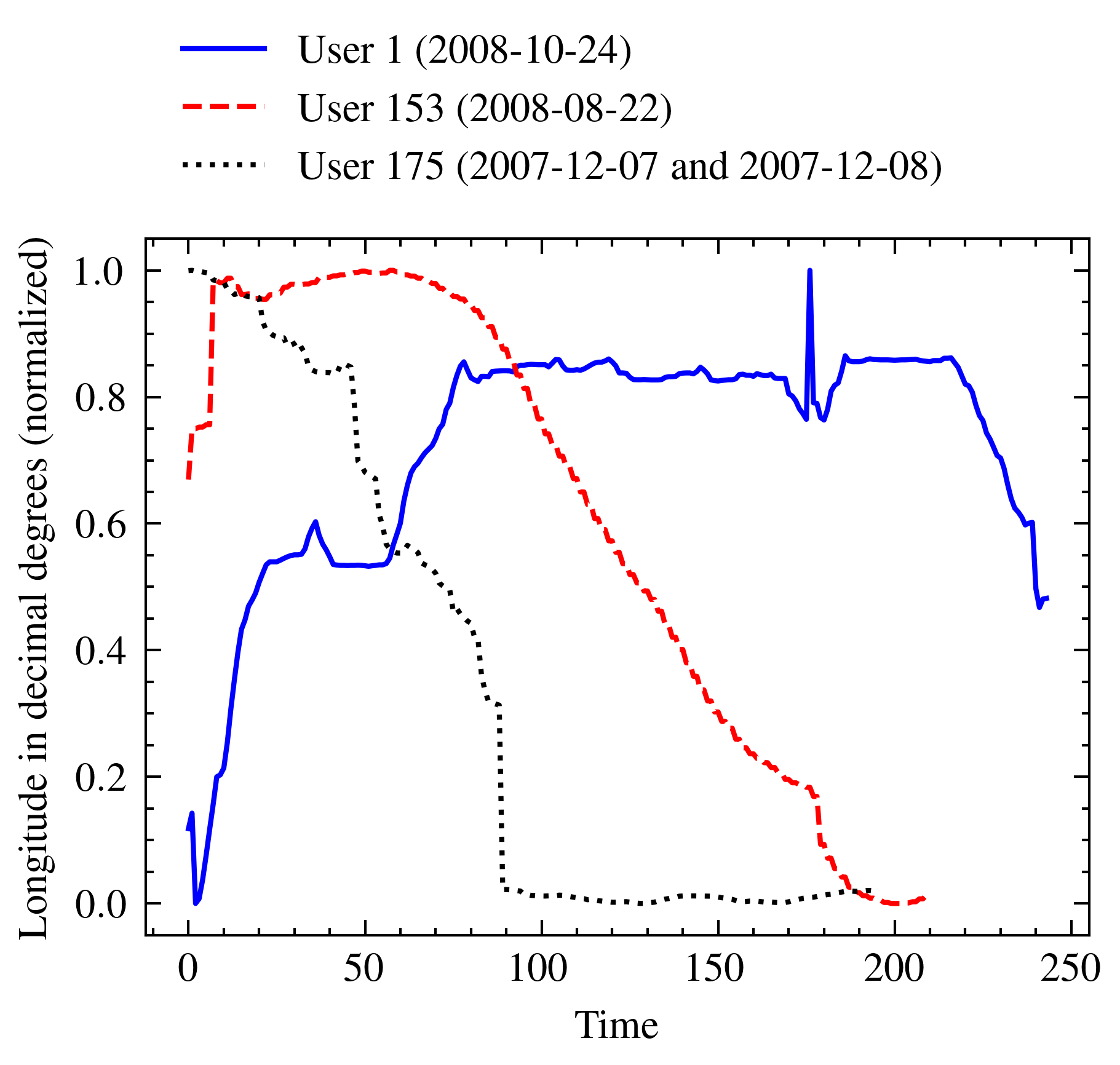}
       \caption{\footnotesize Overview of the longitude variable for the three users.}
       \label{fig:long}
\end{figure}

To keep the prediction procedure simple (especially for the QRC approach), we selected a part of each user's trajectory that contains approximately 200 data points. Specifically, first, for user \#1, we selected the trajectory recorded between 2008-10-24 at 02:09:59 and 2008-10-24 at 02:47:06, which contains 244 data points. Second, for user \#153, we selected the trajectory recorded between 2008-08-22 at 13:10:34 and 2008-08-22 at 13:14:04, which contains 211 data points. Finally, for user \#175, we selected the trajectory recorded between 2007-12-07 at 23:07:44 and 2007-12-08 at 01:27:35, which contains 194 data points. The time series data of theses three users are illustrated in Fig.~\ref{fig:lat} and Fig.~\ref{fig:long} in which the former figure shows the latitude variable of the corresponding user while the latter figure shows its longitude variable. Note that, working with very long user trajectories (thousands of data points) is also interesting but more complex especially using quantum computers. We keep the study of this interesting challenge for our future work. 

We used the Python programming language to conduct our simulations on a MacBook Pro with Apple M1 Pro chip, an 8-core CPU, and a 16 GB RAM. In the sequel, first, we discuss the LSTM and the ESN architectures. Then, we discuss the QRC architecture. Finally, we present the results of the prediction.

\subsection{LSTM Architecture and Parameters}
The Keras library in Python is used to implement the deep learning architecture of LSTM. We used a sequential architecture with three layers, two LSTM layers and one dense layer. The first LSTM layer implements 256 LSTM units with return output sequences set to true (return the last output in the output sequence). The next LSTM layer contains 128 LSTM units with return output sequences set to false (return the full sequence). The last layer is a dense layer and it contains 2 units where the best unit is always selected for prediction. Between each layer there is a dropout of $0.2$ that is added. The tensorflow optimizer RMSprop is used for training with a learning rate of $0.0001$ and the mean-squared error function as the loss function. The training lasted for a period of 2000 epochs where a batch of size 32 is used. The prediction is made for the last 30 timesteps of each time series and the remaining $\ell-30$ are used for training ($\ell$ denotes the length of each time series). The training data are constructed using a sliding window of 70 timesteps, i.e., we built a training data of 70 features.

\subsection{ESN Architecture and Parameters}
We used the pyESN library~\cite{pyesn} in Python. An ESN is created in pyESN with one input unit, one output unit, 500 reservoir units. The spectral radius of the recurrent weight matrix is set to $0.95$ to guarantee the echo-state property. A sparsity of $0.1$ is chosen, which represents the proportion of the recurrent weights set to zero. A noise of $0.001$ is added to each neuron (used for regularization). 
We perform a 2-step prediction for 15 timesteps; for a total length of 30 timesteps.

\begin{table*}
\caption{\small Mean squared error for user \#1 (left), user \# 153 (middle), and user \#175 (right).}
\label{tab1}
\begin{tabular}{|c|c|c|}
\hline
& \multicolumn{2}{|c|}{\textbf{Time Series Variables}} \\
\cline{2-3} 
& \textbf{\textit{Latitude}}& \textbf{\textit{Longitude}}\\
\hline
QRC & $\mathbf{0.000131}$ & $\mathbf{0.001300}$ \\
\hline
ESN & $0.003020$ & $0.002620$  \\
\hline
LSTM & $0.004530$  & $0.007710$ \\
\hline
\end{tabular}
\hfill
\begin{tabular}{|c|c|c|}
\hline
& \multicolumn{2}{|c|}{\textbf{Time Series Variables}} \\
\cline{2-3} 
& \textbf{\textit{Latitude}}& \textbf{\textit{Longitude}}\\
\hline
QRC & $\mathbf{0.000213}$ & $\mathbf{0.000050}$ \\
\hline
ESN & $0.000690$ & $0.000236$ \\
\hline
LSTM & $0.001760$ & $0.003800$ \\
\hline
\end{tabular}
\hfill
\begin{tabular}{|c|c|c|}
\hline
& \multicolumn{2}{|c|}{\textbf{Time Series Variables}} \\
\cline{2-3} 
& \textbf{\textit{Latitude}}& \textbf{\textit{Longitude}}\\
\hline
QRC & $\mathbf{0.000043}$ & $\mathbf{0.000022}$ \\
\hline
ESN & $0.000376$ & $0.000276$ \\
\hline
LSTM & $0.000166$ & $0.00004$ \\
\hline
\end{tabular}
\end{table*} 

\subsection{QRC Architecture and Parameters}
We used the Python library Qiskit~\cite{Qiskit} to implement the QRC framework and we used the Sci-Kit learn library to perform the prediction using the linear regression. We implemented a QRC system with 4 qubits with a magnetic coupling $h=0.5$. The Hamiltonian with a full interaction between the qubits is assumed where every qubit interacts with all other qubits. The inter-qubit interaction coefficients $J_{i,j}$ are chosen randomly using a beta distribution of parameters $\alpha=\beta=0.9$. \if A quantum instance is created using the Qasm simulator and an optimization level of zero is chosen.\fi The time series data are injected into the QRC with a washout period of 70 timesteps to forget the dependence to the initial conditions. The QRC is left evolving according to the Hamiltonian evolution and the gate-based quantum circuits shown in Fig.~\ref{fig:quantum:circuit} are implemented in IBM quantum computers using Qiskit Qasm simulator. The quantum experiment is repeated 1024 times for each entry in the time-series to obtain the average values as outputs of the QRC. A training is then performed to obtain the weight matrix $\mathbf{W}^\text{out}$ using the Moore–Penrose pseudo inverse.
\subsubsection{Notes on the use of real quantum devices}
We also implemented the QRC approach in IBM real quantum computer using 5 qubits. More details are given in the sequel (see subsection~\ref{real}). Note that using quantum computers with more qubits is a promising avenue for solving more interesting and challenging mobility management problems in ITSs that we will study in the future. Our current work will provide useful insights and important future research directions and surely will contribute to the advancement of knowledge. Further, our work can be used to evaluate other complex quantum machine learning solutions.

\subsection{Simulation Results}
First, we calculate, for each method, the mean-squared error (MSE) between the 30-step predicted output and the original data. The MSE of all three methods are illustrated in Table~\ref{tab1}. We can see that the QRC approach offers accurate predictions for different time series variable (longitude vs. latitude) and for different time series data (different users) even though only 4 qubits were used. On the contrary, the RNN approaches (either ESN or LSTM) used lots of neurons (256 in LSTM and 500 in ESN) to only produce inferior results compared to the QRC approach. One of the main ideas behind using QRC was to take advantage of its capacity to produce a complex dynamical system through the Hamiltonian evolution and the high-dimensional complex Hilbert space. Table~\ref{tab1} shows that indeed quantum computing can be used to predict, with low MSE, complex time series data. Note, however, that the application of the QRC approach to more complicated time series (longer or multivariate) is important and should be analyzed in our future work. It will be probably evident that for more complex time series data, the number of qubits should be increased and thus the trade-off between performance in terms of prediction and complexity in terms of time should be analyzed carefully for the QRC approach.

In the next figures, we visualize the 30-step predictions produced by each method; QRC, ESN, and LSTM.

\begin{figure}[!b]
    \centering
       \includegraphics[scale=1]{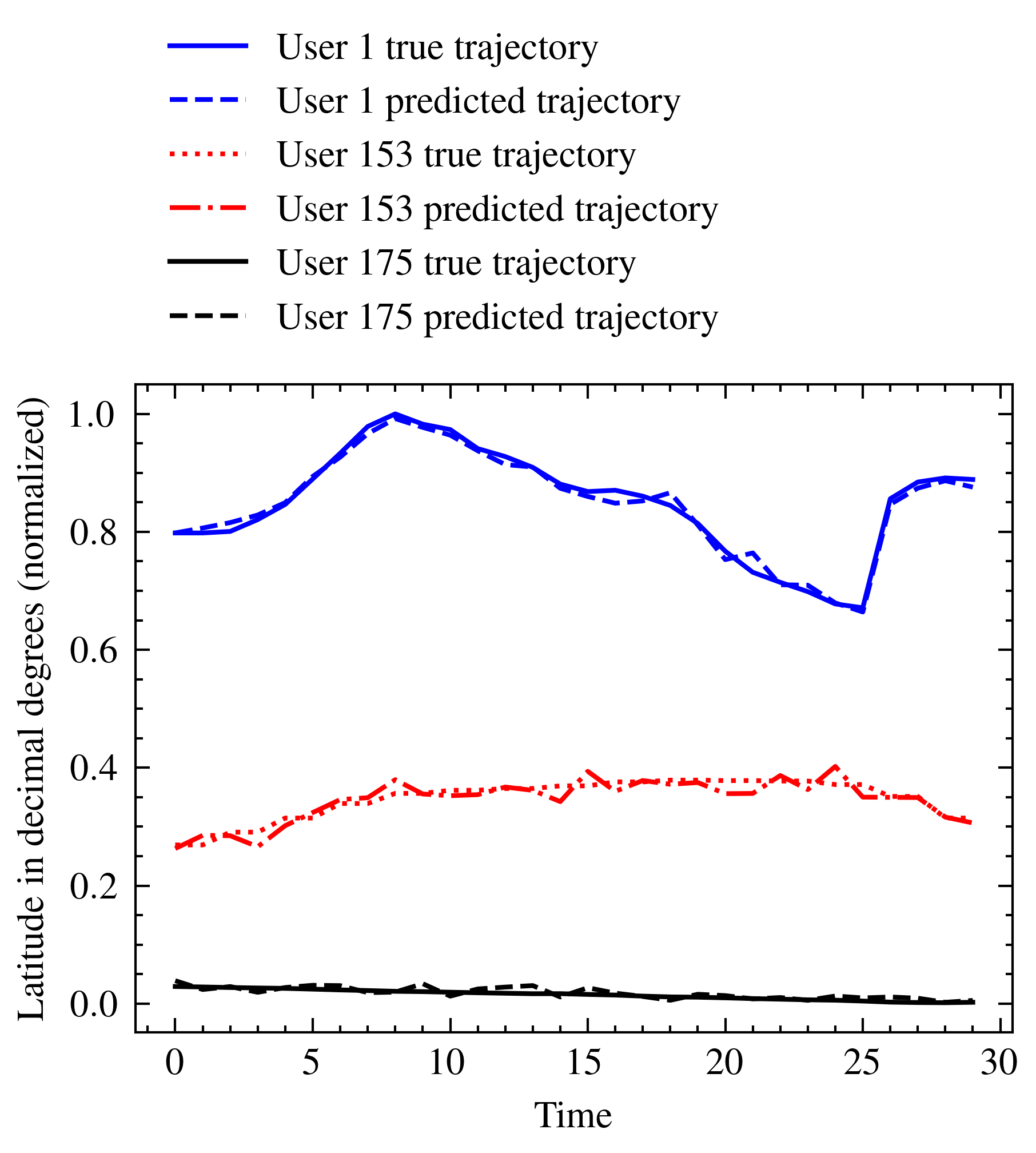}
       \caption{\footnotesize QRC $30$-steps prediction of the latitude variable for the three users.}
       \label{fig:lat:qrc:users}
\end{figure}
Fig.~\ref{fig:lat:qrc:users} illustrates the prediction produced by the QRC approach on the time series data of the three users for the latitude variable. We can see that the prediction is very close to the true trajectory data. \if and this is true for the latitude variable as well as the longitude variable.  The degrees of freedom of the QRC help in producing a system that is able emulate nonlinear dynamical system.\fi

\begin{figure}[!h]
    \centering
       \includegraphics[scale=1]{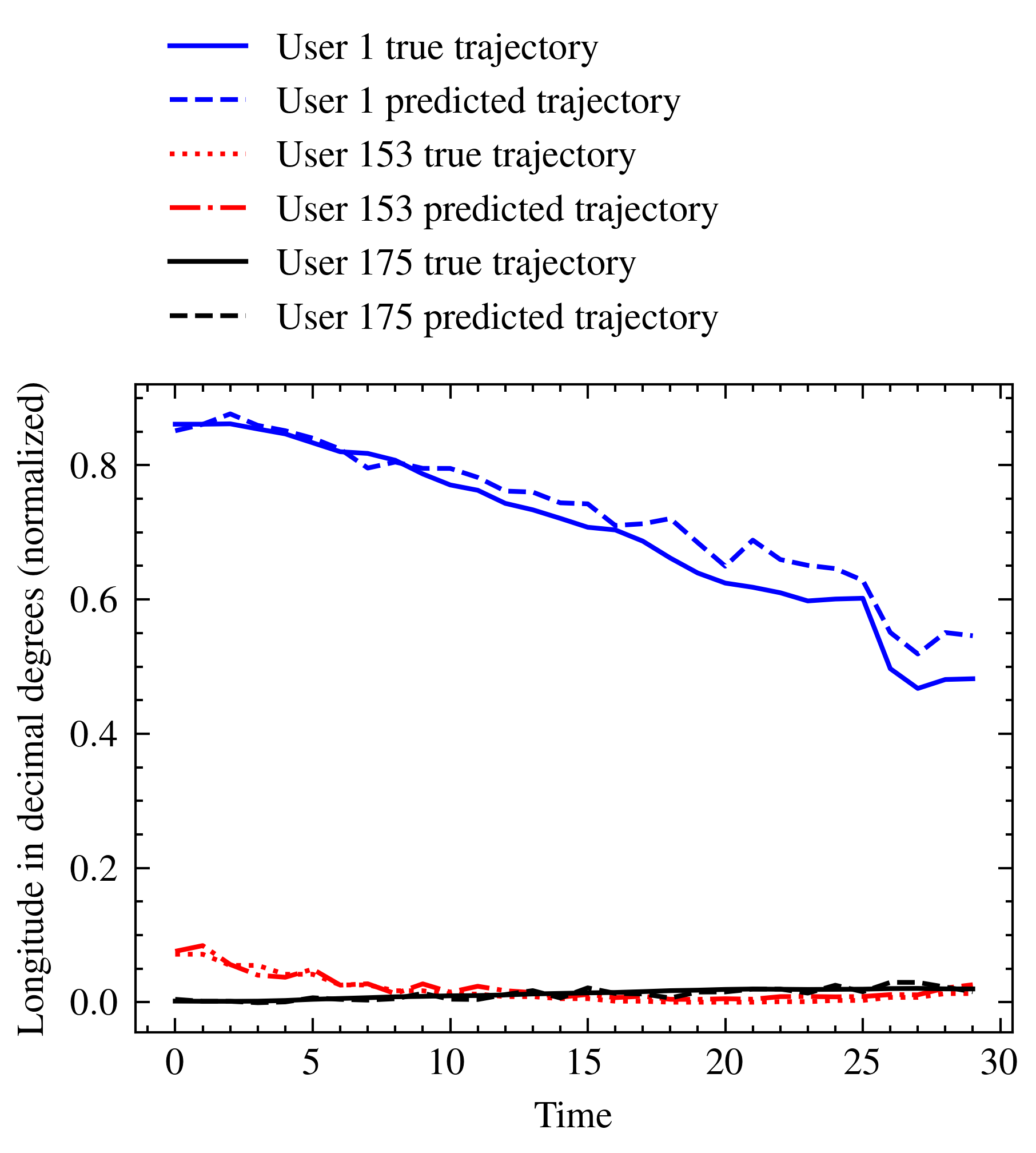}
       \caption{\footnotesize QRC $30$-steps prediction of the longitude time series variable for the three users.}
       \label{fig:long:qrc:users}
\end{figure}
Fig.~\ref{fig:long:qrc:users} shows a visualization of the prediction produced by the QRC approach on the time series data of the three users for the longitude variable. The QRC prediction is again very accurate as the predicted signal is very close to that of the true data. Despite the change of the time series variable, the QRC is able to produce a good prediction for all users, which illustrate the remarkable feature of the QRC system. When the time series is similar to a straight line (the prediction for user \#175 in Fig.~\ref{fig:long:qrc:users}), the prediction might not look smooth and perfect but the MSE is still very low. For this particular user, its original time series is almost constant especially for the last 100 timesteps, which is caused by an overfitting problem.

\begin{figure}[!b]
    \centering
       \includegraphics[scale=1]{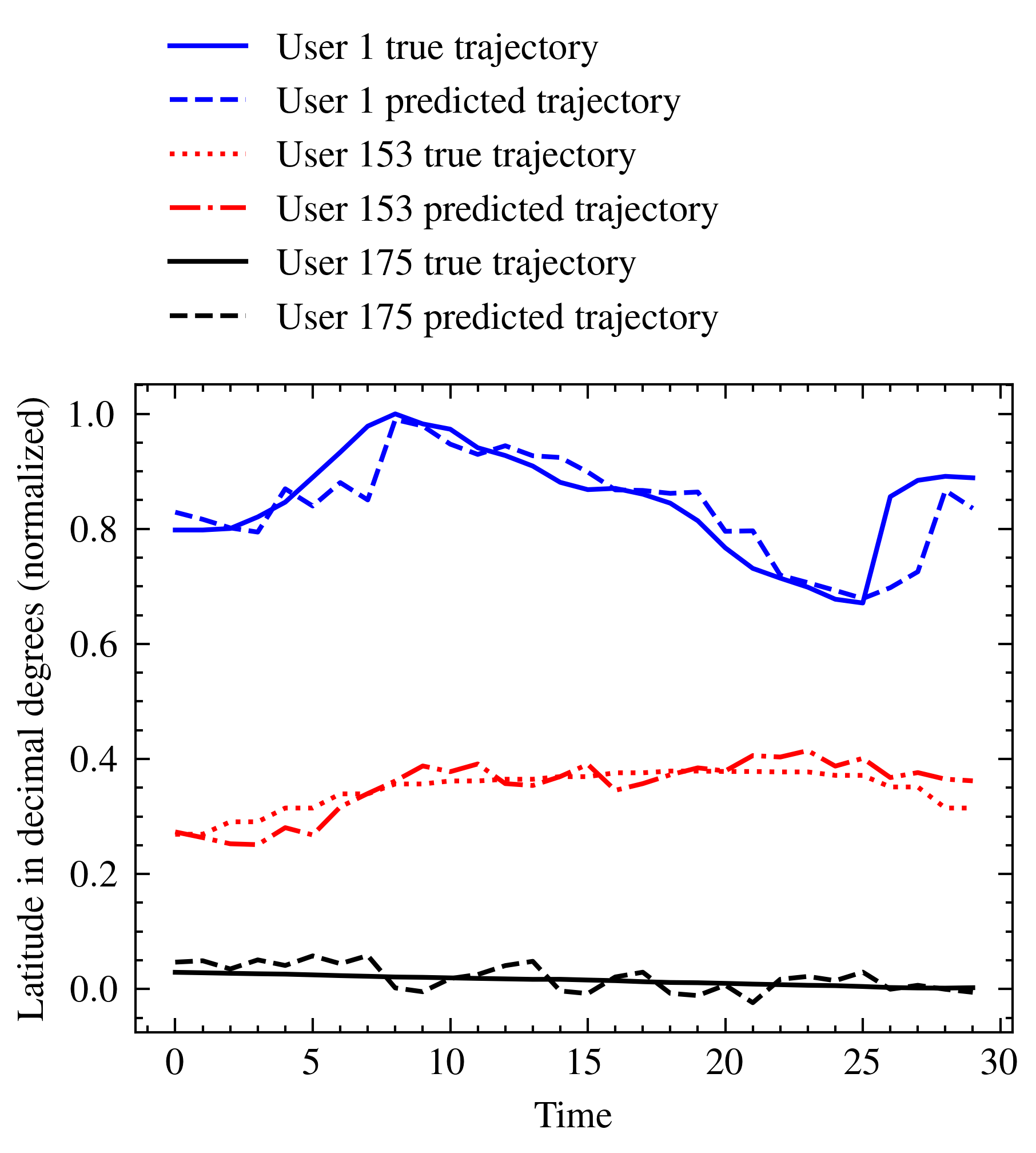}
       \caption{\footnotesize ESN $30$-steps prediction of the latitude time series variable for the three users.}
       \label{fig:lat:esn:users}
\end{figure}
\begin{figure}[!h]
    \centering
       \includegraphics[scale=1]{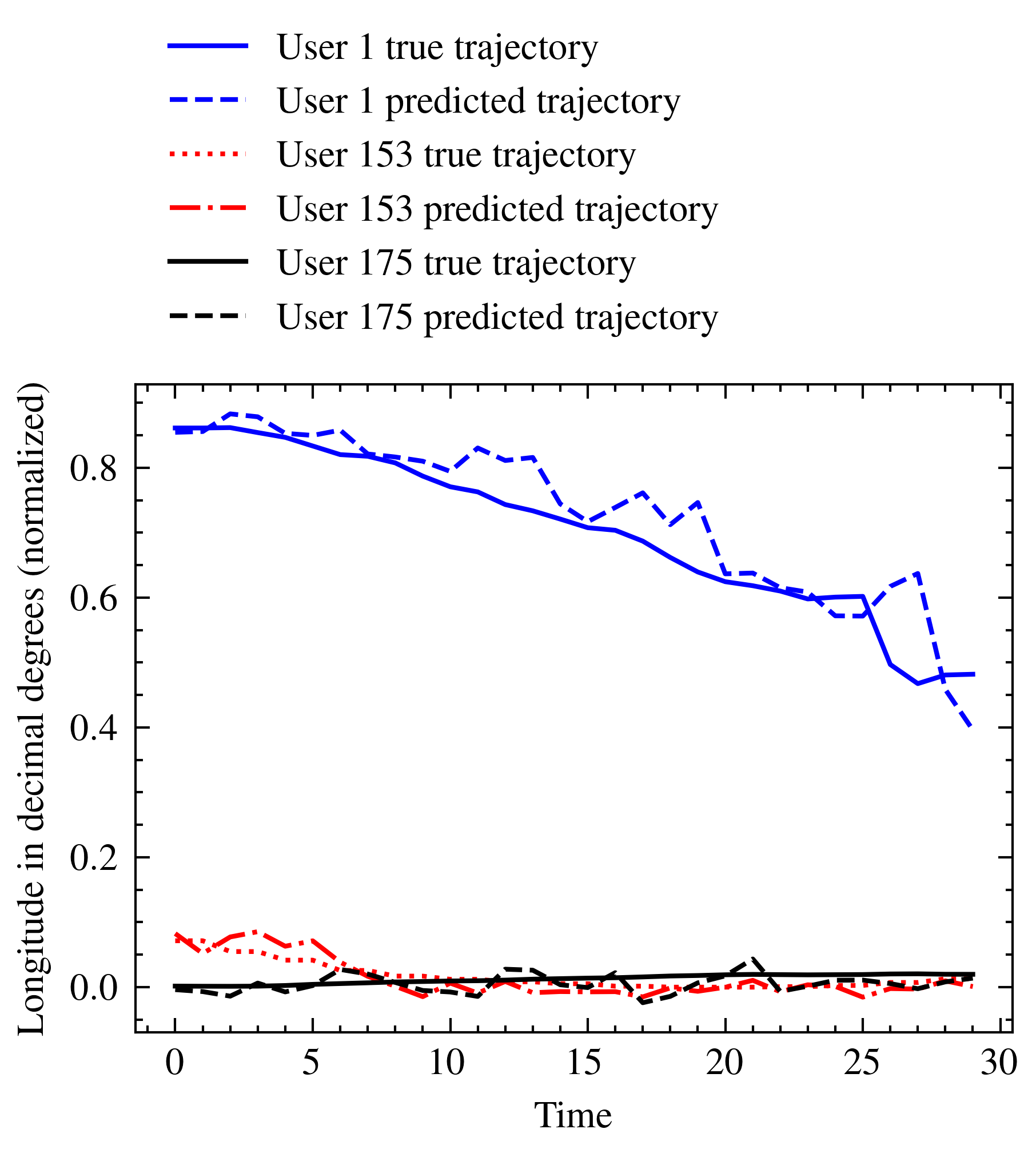}
       \caption{\footnotesize ESN $30$-steps prediction of the longitude time series variable for the three users.}
       \label{fig:long:esn:users}
\end{figure}

In Fig.~\ref{fig:lat:esn:users} and Fig.~\ref{fig:long:esn:users} we visualize the prediction produced by the ESN approach on the time series data of the three users for the latitude variable and the longitude variable respectively. The ESN approach produces also good prediction as the predicted signals and their corresponding true values are close to each other. Further, the ESN approach produces similar predictions compared to the QRC approach. Nonetheless, the latter is better and has lower MSE as well. It is also true for the ESN approach, as in the QRC approach, that the prediction for the user \# 175 is more difficult than the remaining users.

\begin{figure}[!h]
    \centering
       \includegraphics[scale=1]{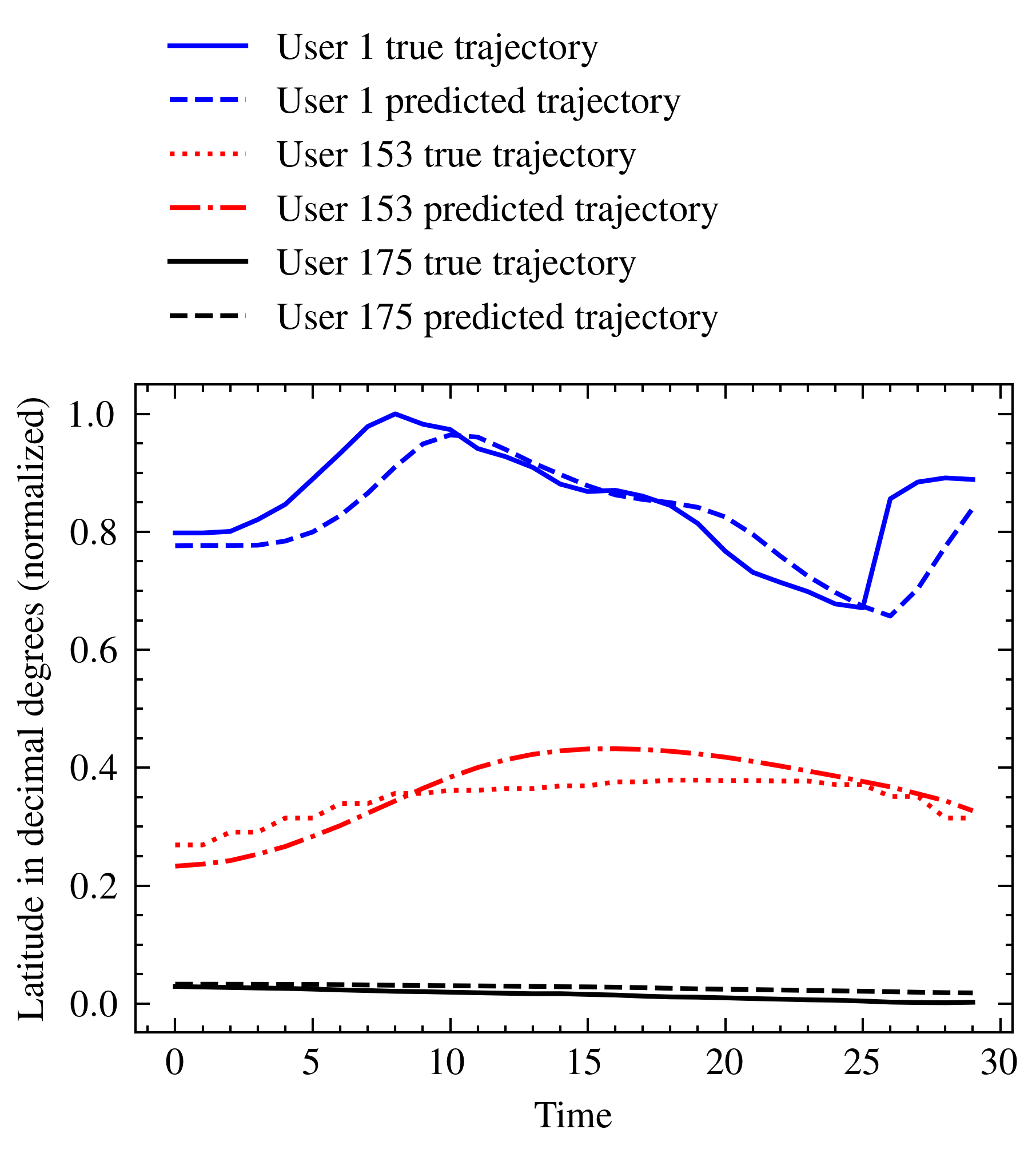}
       \caption{\footnotesize LSTM $30$-steps prediction of the latitude time series variable for the three users.}
       \label{fig:lat:lstm:users}
\end{figure}

\begin{figure}[!h]
    \centering
       \includegraphics[scale=1]{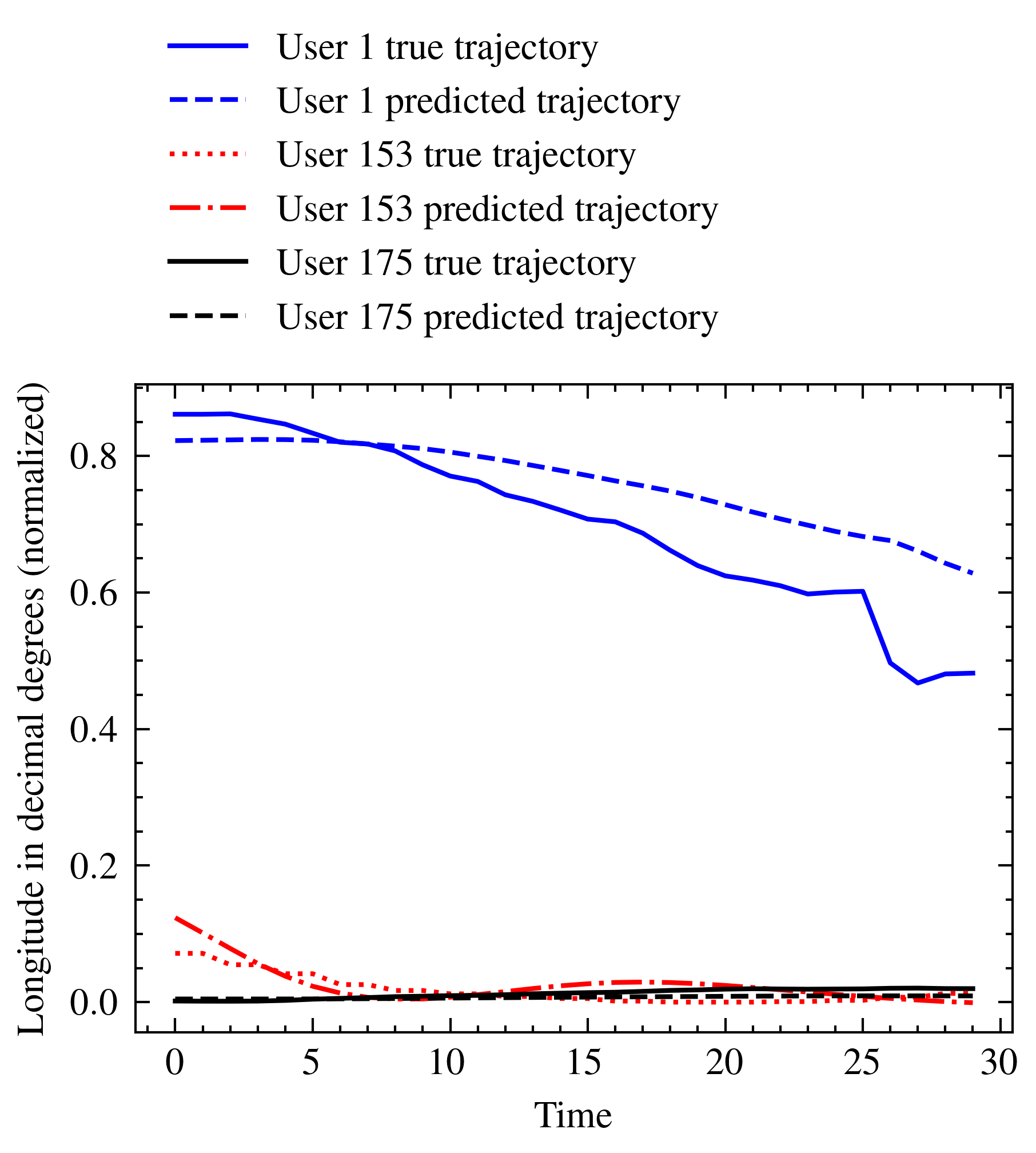}
       \caption{\footnotesize LSTM $30$-steps prediction of the longitude time series variable for the three users.}
       \label{fig:long:lstm:users}
\end{figure}
In Fig.~\ref{fig:lat:lstm:users} and Fig.~\ref{fig:long:lstm:users} we visualize the prediction produced by the LSTM approach on the time series data of the three users for the latitude variable and the longitude variable respectively. It is clear that the LSTM prediction is the worst among all approaches. This was also abserved in Table~\ref{tab1} where the MSE was the worst for the LSTM prediction for all users (except for user \#175).

\subsection{Real Quantum Computer Implementation Results}\label{real}
In this part, we obtain the QRC prediction of the longitude time series variable for  user \#1 with a real quantum computer and we compared the result to the ESN and LSTM predictions. We access IBM quantum computers through the provider \texttt{ibm-q/open/main/} that provides access to six real quantum computers. We implement the QRC approach on the version 1.1.34 of the \texttt{ibmq-quito} 5-qubits quantum computer that is equipped with a processor of type Falcon r4T. Due to limited available resources on IBM quantum computers, i.e., only a maximum of 20000 shots and a maximum of 100 circuits are permitted, we selected a small-size time series to implement the QRC approach. We selected the longitude of the user \#1 recorded between 2008-10-27 at 11:54:49 and 2008-10-27 at 12:05:54, which contains 50 data points. We used 10 timesteps for the washout period and we performed a 5-timesteps prediction. Due to the natural quantum noise and gate errors, we increased the number of shots to $4000$ and we increased the optimization level to 3 to optimize the quantum circuits at the expense of longer transpilation time.

\begin{figure}[!h]
    \centering
       \includegraphics[scale=1]{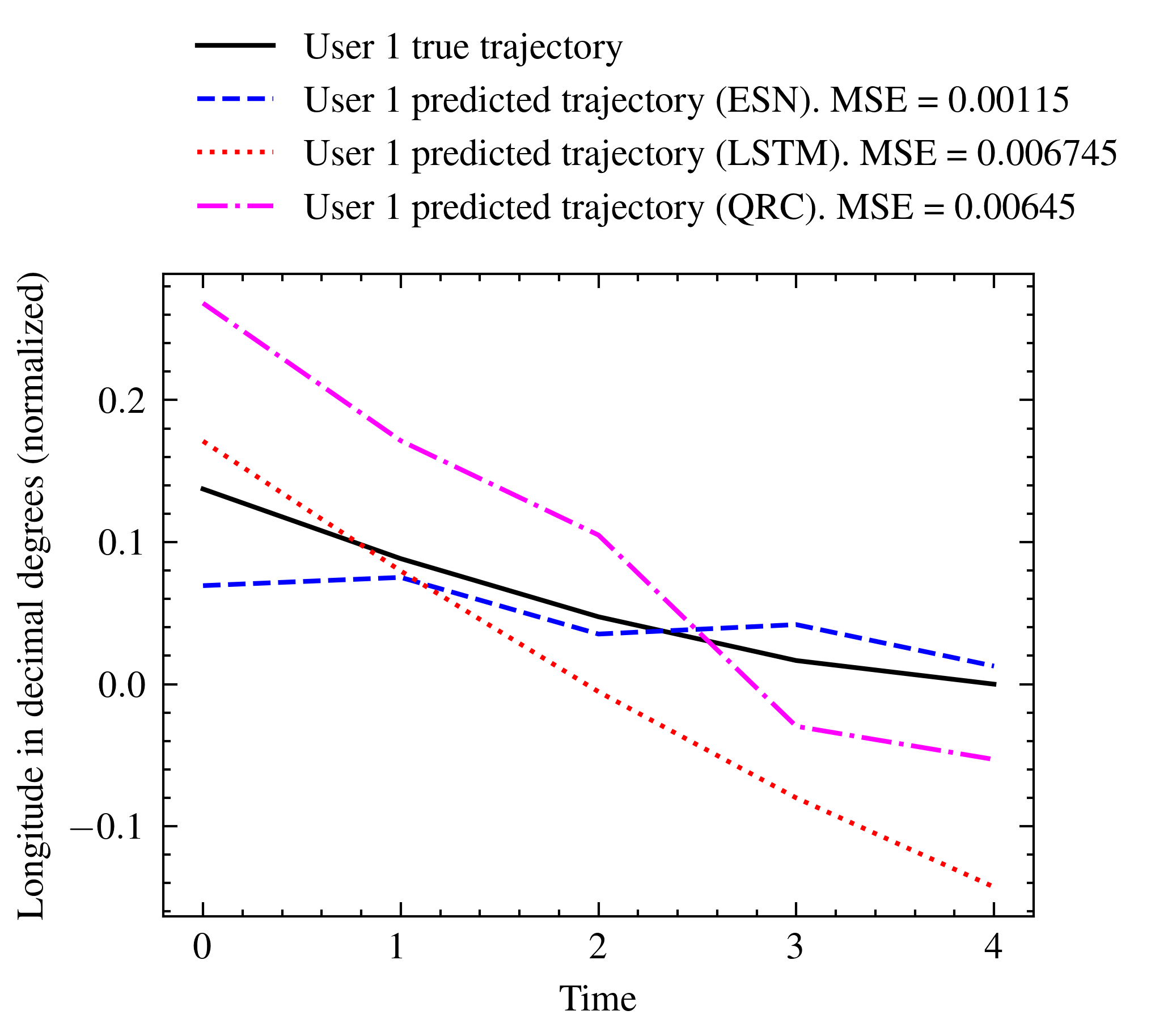}
       \caption{\footnotesize Real quantum computer $5$-steps prediction of the longitude time series variable for user \#1.}
       \label{fig:long:qrc:esn:lstm}
\end{figure}
We observe from Fig.~\ref{fig:long:qrc:esn:lstm} that the LSTM prediction produces very close performance to the QRC prediction but the latter is still better. Even though current real quantum computers are not fault-tolerant, we demonstrated that the QRC prediction approach is able to produce satisfactory results in terms of MSE compared to the well-known and most used deep learning-based prediction approaches such as the LSTM and the ESN approaches. Note that the prediction is done on small-size time series and a more sophisticated quantum prediction approach might be needed for more complex time series. Despite this, the QRC prediction approach is appealing and worth considering in real world scenarios.

\subsection{Discussions}
We observed through our analysis of the different results that implementing the QRC approach on open-access IBM quantum experience is easy especially for small-size time series and when the number of qubits is not large. However, for large-size time series or when the number of qubits is large, one needs more resources in terms of time and computation. We observed that more advanced quantum computers can forecast efficiently more challenging and interesting time series machine learning problems compared to their classical counterparts. Also, the QRC approach is appealing to solve wireless networks machine learning tasks such as the prediction and classification of time series, e.g., the prediction of wireless traffic, the prediction of signal-related measurements, etc. 


\section{Conclusions}\label{cl}
In this paper, we applied a quantum reservoir computing (QRC) approach, proposed recently to predict time series data, to the user mobility prediction problem in mobile wireless networks. We used a real-world GPS dataset with hundred timesteps. The QRC approach exploited the rich dynamics of quantum mechanics to produce a highly complex and dynamical quantum system capable of producing highly accurate prediction of GPS trajectories. The QRC approach was compared to two classical approaches; the long-short term memory (LSTM) and the echo-state network (ESN). Both classical approaches are based on the recurrent neural network framework. The ESN uses the reservoir computing technique similar to the quantum one. The difference is that the quantum technique constructs a quantum reservoir based on the quantum mechanics whereas the ESN constructs a classical reservoir based on random matrix generation. The quantum reservoir used the time evolution of the fully connected transverse field Ising Hamiltonian and we showed how we can produce this time evolution using quantum simulation with gate-based quantum circuits. We showed that the QRC approach is able to produce the lowest prediction error in terms of mean-squared error compared to the LSTM and the ESN approaches. 

In the future, it is worth investigating the QRC approach for multivariate time series that have much longer length. It is also interesting to investigate the trade-off between complexity and performance of the QRC for higher number of qubits. Finally, hyper-parameters optimization should be carefully done for the QRC approach as well as for other classical approaches to understand in depth what parameters influence the performance of predicting time series.

\section*{Acknowledgment}
The authors would like to thank the Fonds de recherche du Québec - Nature et technologies (FRQNT) as well as the Natural Sciences and Engineering Research Council of Canada (NSERC) for their financial supports.

\bibliographystyle{IEEEtran}
\bibliography{IEEEabrv,refs}

\end{document}